# Fabrication and transport properties of $Sr_{0.6}K_{0.4}Fe_2As_2$ multifilamentary superconducting wires

Chao Yao, Yanwei Ma[*], Chengduo Wang, Xianping Zhang, Dongliang Wang, Chunlei Wang,

He Lin, Qianjun Zhang

Key Laboratory of Applied Superconductivity, Institute of Electrical Engineering,

Chinese Academy of Sciences, PO Box 2703, Beijing 100190, China

[*]E-mail: ywma@mail.iee.ac.cn

**Abstract**

Seven-core Ag/Fe sheathed $Sr_{0.6}K_{0.4}Fe_2As_2$ (Sr-122) superconducting wires were produced by the *ex situ* powder-in-tube (PIT) method. The relationship between the cold-work deformation process and the superconducting properties of wires were systematically studied. It was found that flat rolling can efficiently increase the density of the superconducting core and largely improve the transport critical current density ($J_c$) of as-drawn wires. The $J_c$ of the best sample achieved 21.1 kA/cm$^2$ at 4.2 K in self field, and showed very weak magnetic field dependence in high fields. Our result suggested a promising future of multifilamentary iron-based superconductors in practical application.

___________________________________________________

[*] Author to whom any correspondence should be addressed.

Among several types of iron-based superconductors discovered in 2008 [1-6], the 122 type ($AeFe_2As_2$, Ae = alkali or alkali earth elements) shows relatively high superconducting transition temperatures ($T_c$) up to 38 K [3], very high upper critical fields ($H_{c2}$) above 100 T [7,8] with small anisotropy (γ = $H_{c2}//$ab/ $H_{c2}//$c) about 1.5-2 [9,10], and large critical current density $J_c$ over $10^6$ Acm$^{-2}$ in thin films [11-13], suggesting very great potential in high-field applications. As an important role in practical applications such as cables and magnets, high-performance superconducting wires need to be developed. On the other hand, since the core of wire is polycrystalline, the transport $J_c$ of iron-based superconducting wire is largely deteriorated by high-angle grain boundaries [14, 15] and defects such as cracks, pores, and impurities [16-19]. In recent years, the grain connectivity and weak-linked grains was improved by chemical doping [20, 21] and rolling induced c-axis texture of grains [22, 23], respectively. Recently, the transport $J_c$ of Sr-122 tapes with Sn addition and c-axis texture have been greatly increased up to 15 kA/cm$^2$ at 10 T and 4.2 K by Gao et al [24]. By HIP technique and low-temperature sintering to obtain high density and fine grains, Weiss et al also achieved a high transport $J_c$ of 100 kA/cm$^2$ in self field and 8.5 kA/cm$^2$ at 10 T in Ba-122 wires [25].

Towards practical applications of iron-based superconductors, fabricating multifilamentary wires and tapes is an indispensable step. Three-core and seven-core FeSe wires have been realized until now, but their properties such as $J_c$ and $T_c$ are very limited [26, 27]. On the other hand, as mentioned above, it is significant that the transport $J_c$ of 122 single-core wires has been greatly enhanced these years, so the 122 compounds are the most promising candidate for fabricating multifilamentary wires. To date, there is still no report on the multifilamentary wires of 122 type iron-based superconductors. In this work, we first prepared seven-core Ag/Fe sheathed Sr-122 wires with powder-in-tube (PIT) method, which was widely used in fabricating single-core iron-based superconducting wires [28], as well as multifilamentary bismuth-based [29, 30] and $MgB_2$ [31, 32] wires. In order to enhance the homogeneity and grain connectivity of material, $Sr_{0.6}K_{0.4}Fe_2As_2$ was synthesized with *ex situ* method and added with proper Sn. Two deformation routes were used in cold-work process of the multifilamentary wires, one was just drawing to get wires with various diameters, and the other was rolling the as-drawn wires into tapes with different thickness. The relationship between the cold-work deformation and the superconducting properties of the samples were systematically studied.

The starting materials for synthesizing $Sr_{0.6}K_{0.4}Fe_2As_2$ precursor are small Sr pieces, K bulks and fine Fe and As powders. In order to compensate the loss in the sintering, 20% K was extra added [33]. They were mixed and ground with ball milling in Ar atmosphere for about 12 h, then sealed and sintered in an Nb tube for 35 h at 900 $^o$C. The precursor was added with 10 wt. % Sn and ground into fine powder in an agate mortar in Ar atmosphere. The final powder was packed into Ag tubes (OD: 8 mm and ID: 5 mm), which were drawn into wires of 1.58 mm in diameter with a reduction rate of about 5 % and cut into short wires. As shown in figure 1, seven short Ag sheathed wires were then bundled into a Fe tube (OD: 8 mm and ID: 5 mm), and drawn into wires of about 2.00, 1.90, 1.80 and 1.70 mm in diameter with a reduction rate of about 5 %. Some multifilamentary wires of 2.00 mm in diameter were further rolled into tapes of about 1.50, 1.25, 1.00, 0.80 and 0.60 mm in thickness. This cold-work process is also shown in figure 1. The Ag inner sheath of these multifilamentary wires can avoid nonsuperconducting layer produced by the reaction between sheath material and Sr-122 core [20], but the temperature for final heat treatment should be below the melt point of Ag (~ 961 $^{o}$C) . Therefore, the wires and tapes were sealed into quartz tubes, and heat treated at 850 ~ 950$^{o}$C for about 0.5 h. For the x-ray diffraction (XRD) characterization and comparison the transport property with seven-core tapes, Fe sheathed single-core PIT Sr-122 tapes were prepare with the same precursor, Fe tubes, packing density, cold-work process and heat treatment as above.

The resistance measurements were carried out on a Quantum Design's physical property measurement system (PPMS) using the four-probe method. The microstructure of Sr-122 cores was analyzed with a Hitachi S4800 scanning electron microscope (SEM). The area of cross sections and Sr-122 cores of the samples was determined from the corresponding SEM images. The element distribution on the cross section and the actual composition of the Sr-122 cores were examined with an energy dispersive x-ray spectroscopy (EDX). The phases of the superconducting core were studied by the x-ray diffraction (XRD) on a Rigaku D/MAX 2500 diffractometer. The transport critical current $I_c$ was measured at 4.2 K using short wire and tape samples of 3 cm in length with standard four-probe method and evaluated by the criterion of 1 $\mu$V/cm, and the critical current was divided by the cross section area of the superconducting cores to get the critical current density $J_c$.

The SEM images in figure 1 shows the cross section of well-developed seven-core Sr-122 wires and tapes, even the as-drawn wires are rolled into tapes of 0.6 mm thick, the Ag inner sheath can still separate and clad the seven Sr-122 superconducting cores very well. The nominal diameter or thickness, area of the whole cross section ($S_{Fe+Ag+Sr122}$) and Sr-122 cores ($S_{Sr122}$), on set superconducting transition temperature ($T_{c,\ onset}$), resistivity at 300 K ($\rho_{300}$), and transport $J_c$ of each sample are list in table 1. To investigate the phases of Sr-122 superconducting core, after peeling away the Fe sheath, the Sr-122 core was ground to get the powder sample for the XRD characterization, and the result is shown in figure 2. Except for some minor peaks of impurities and added Sn, the XRD pattern shows strong peaks of Sr-122 phase, which are indexed in the figure. Figure 3 (b) is an optical image taken by a microscope at the area marked by the arrow in figure 3 (a), showing one of the seven cores of tape (thickness = 0.60 mm). It can be seen that the Sr-122 core looks very dense, and the boundary between the core and Ag inner sheath is very clear, indicating that there is no reaction layer after heat treatment. We examined the area marked by the white rectangular frame in figure 3 (a) with EDX. Comparing the original image taken by SEM in figure 3 (c) with the corresponding EDX element mapping in figure 3 (d), it is obvious that the areas of Sr-122, Ag sheath and Fe sheath in these two images are well accordant and tightly connected. In order to determine the actual composition of the Sr-122 cores, two of the cores were also analyzed with EDX, and the selected areas were marked with white rectangular frames in figure 3 (a). The result of area 1 and 2 is $Sr_{0.59}K_{0.41}Fe_2As_{1.99}$ and $Sr_{0.59}K_{0.37}Fe_2As_{2.04}$ respectively. Though the amount of K is a little inhomogeneous, this result is generally close to the nominal composition. The results above suggest that the Ag/Fe sheath is suitable for fabricating Sr-122 multifilamentary wires and tapes.

Figure 4 (a) and (b) shows the temperature dependence of the resistivity for the wires and tapes respectively. Here the resistivity $\rho$ can be define by $S_{Fe+Ag+Sr122}/\rho = S_{Fe}/\rho_{Fe} + S_{Ag}/\rho_{Ag} + S_{Sr122}/\rho_{Sr122}$. Since the resistivity of $Sr_xK_{1-x}Fe_2As_2$ single crystal [4] is much larger than that of Ag and Fe at 300 K, and the $S_{Sr122}$ is just about 8% ~ 9% of $S_{Fe+Ag+Sr122}$, the value of $\rho$ is mainly determined by the Ag and Fe sheath at this temperature. As list in table 1, the resistivity at 300 K for both wires and tapes decrease with the reducing of $S_{Fe+Ag+Sr122}$ and $S_{Sr-122}$. Therefore, the contact resistance between the Ag and Fe sheath should be also taken into consideration. With the increase of deformation force, the contact between the Ag and Fe sheath gets more tight, which will reduce the resistance and benefit the transport property. For the wires with different diameters, the $T_{c,\,onset}$ is almost the same during the deformation. On the other hand, the $T_{c,\,onset}$ of tapes first decrease a little after rolled to 1.50 mm thick, then seem to recover after further rolled to 1.25 and 1.00 mm thick, and finally increase for about 1 K in tapes of 0.80 and 0.60 mm in thickness.

Referring to table 1, it can be noted that this change of $T_{c,\,onset}$ is very accordance with that of transport $J_c$ at 4.2 K in self field, which can be ascribe to the change of density caused by deformation. It is konwn that the rolling can enhance the density of the superconducting core, and thus improve the grian connectivity, which is very critical to the tansport propety. On the other hand, we can also conclude that the drawing in our work does not change the density of Sr-122 core much. For the rolling samples, the homongenous core after drawing is somewhat damaged during transition from round wire to rectangular tape at the beginning, then recover and become even more dense by subsequent rolling. The increase of density can also improve the reaction and formation of Sr-122 phase, achieve a higher $T_{c,\,onset}$. Compared to the wires, the tansition width of the tapes is mone boarder, indicating that the density become inhomogenous in Sr-122 cores and different among cores at the different place of the tapes, like the case of Bi-2223 multifilamentary tapes [35].

Figure 5 presents the field dependent transport $J_c$ of tapes rolled to 1.00, 0.80 and 0.60 mm in thickness, and the $J_c$ also significantly increase with the reducing of the rolling thickness in high fields. The $J_c$ of 0.60 mm thick tape is the highest, which achieves 21.1 kA/cm$^2$ at 4.2 K in self field and 3.3 kA/cm$^2$ at 10 T. It decreases with the increasing field to about 0.5 T, and then shows very weak field dependence up to 10 T. For comparison, the transport $J_c$ of the single-core tape of 0.60 mm thick is also shown in the figure (see the dashed line). We can see that prepared with the same packing density and rolled into the same thickness, the transport $J_c$ of seven-core tape is even higher than the single-core tape. Figure 6 (a) and (b) show the SEM image of the cross section of the single-core and seven-core tapes respectivly. It is obvious that the microstructure of the later one is much more denser, which can explain its superior transport $J_c$ value.

In the fabricating of mono- and multifilamentary Bi-2223 tapes, Bi-2212 tapes and MgB$_2$ tapes [34-39], the rolling induced increase of density plays an important role in the improvement of transport current, which is also the same case in Sr-122 tapes in our work. As shown in table 1, for wires of 1.80 mm in diameter and tapes of 0.60 mm in thickness, which are respectively drawn and rolled from wires of 2.00 mm in diameter, the values of deformation rate are almost the same. However, the changes on the transport $J_c$ of Sr-122 cores are different. Drawing has no obvious effect on $J_c$, but rolling can increase the core density and significantly improve the $J_c$ in the whole field region. One the other hand, in figure 6 (b) there are still a few pores and cracks in the superconducting core. Therefore, by optimizing process parameters to obtain denser microstructures, it is possible to further improve the $J_c$ in the future. Moreover, considering the extremely weak field dependence of its $J_c$, the multifilamentary wires and tapes of 122 type iron-based superconductors will be very promising candidates in high-field applications.

In summary, we have first fabricated Fe/Ag sheathed Sr-122 seven-core superconducting wires and tapes by the *ex situ* PIT method. SEM and EDX examination confirmed that the Ag/Fe sheath is very suitable to develop multifliamnerary 122 type wires and tapes. Compared to the drawing deformation, rolling can increase the density of superconducting cores. The transport $J_c$ increases with the reducing of the rolling thickness. By rolling the tape to 0.6 mm thick, we achieved a transport $J_c$ of 21.1 kA/ cm$^2$ at 4.2 K in self field and 3.3 kA/ cm$^2$ at 10 T, which exhibited very weak field dependence in high fields. This value is also higher than the single-core tape with the same rolling thickness. Therefore, rolling can achieve large transport $J_c$ in multifilamentary 122 type iron-based superconducting wires, which suggests a promising future in practical applications.

The authors thank Prof. S. Awaji and Prof. K. Watanabe at the IMR, Tohoku University, Japan for the high field transport current measurements. This work is partially supported by the National '973' Program (grant No. 2011CBA00105) and the National Natural Science Foundation of China (grant Nos. 51025726 and 51172230).

# References

[1] Y. Kamihara, T. Watanabe, M. Hirano and H. Hosono, J. Am. Chem. Soc. **130**, 3296 (2008)

[2] Z. A. Ren, W. Lu, J. Yang, W. Yi, X. L. Shen, C. Z. Li, G. C. Che, X. L. Dong, L. L. Sun, F. Zhou and Z. X. Zhao, Chin Phys. Lett. **25**, 2215 (2008)

[3] M. Rotter, M. Tegel and D. Johrendt, Phys. Rev. Lett. **101**, 107006 (2008)

[4] K. Sasmal, B. Lv, B. Lorenz, A. Guloy, F. Chen, Y. Xue and C. W. Chu, Phys. Rev. Lett. **101**, 107007 (2008)

[5] X. C. Wang, Q. Q. Liu, Y. X. Lv, W. B. Gao, L. X. Yang, R. C. Yu, F. Y. Li and C. Q. Jin, Solid State Commun. **148**, 538 (2008)

[6] F. C. Hsu, J. Y. Luo, K. W. Yeh, T. K. Chen, T. W. Huang, P. M. Wu, Y. C. Lee, Y. L. Huang, Y. Y. Chu, D. C. Yan and M. K. Wu, Proc. Natl Acad. Sci. **105**, 14262 (2008)

[7] N. Ni, S. L. Bud'ko, A. Kreyssig, S. Nandi, G. E. Rustan, A. I. Goldman, S. Gupta, J. D. Corbett, A. Kracher and P. C. Canfield, Phys. Rev. B **78**, 014507 (2008)

[8] X. L. Wang, S. R. Ghorbani, S. I. Lee, S. X. Dou, C. T. Lin, T. H. Johansen, K. H. Müller, Z. X. Cheng, G. Peleckis, M. Shabazi, A. J. Qviller, V. V. Yurchenko, G. L. Sun and D. L. Sun, Phys. Rev. B **82**, 024525 (2010)

[9] A. Yamamoto, J. Jaroszynski, C. Tarantini, L. Balicas, J. Jiang, A. Gurevich, D. C. Larbalestier, R. Jin, A. S. Sefat, M. A. McGuire, B. C. Sales, D. K. Christen and D. Mandrus, Appl. Phys. Lett. **94**, 062511 (2009)

[10] H. Q. Yuan, J. Singleton, F. F. Balakirev, S. A. Baily, C. G. F hen, J. L. Luo, and N. L. Wang, Nature **457**, 565 (2009)

[11] S. Lee, J. Jiang, Y. Zhang, C. W. Bark, J. D. Weiss, C. Tarantini, C. T. Nelson, H. W. Jang, C. M. Folkman, S. H. Baek, A. Polyyanskii, D. Abraimov, A. Yamamoto, J. W. Park, X. Q. Pan, E. E. Hellstrom, D. C. Larbalestier and C. B. Eom, Nat. Mater. **9**, 397 (2010)

[12] T. Katase, H. Hiramatsu, T. Kamiya, and H. Hosono, Appl. Phys. Express **3**, 063101 (2010)

[13] T. Katase, H. Hiramatsu, V. Matias, C. Sheehan, Y. Ishimaru, T. Kamiya, K. Tanabe and H. Hosono, Appl. Phys. Lett. **98**, 242510 (2011)
9

**Captions**

Figure 1 The fabricating process of multifilamentary Sr-122 wires with various diameters and tapes with various thicknesses. SEM image of the cross section of wires (diameter = 2.00 and 1.70 mm) and tapes (thickness = 1.25 and 0.60 mm) are shown.

Figure 2 XRD patterns of powder sample from the single-core Sr-122 tapes. The main peaks of Sr-122 phase are indexed, and peaks of impurities and Sn are marked with the corresponding symbols.

Figure 3 (a) SEM image of the cross section of seven-core Sr-122 tape (thickness = 0.60 mm); (b) Optical microscope image of one of the seven cores (marked by the arrow in (a)); (c) SEM image of the selected area (marked by the white rectangular frame in (a)) for EDX examination and (d) the corresponding EDX mapping image.

Figure 4 Temperature dependence of resistivity for the (a) wires with various diameters and (b) tapes with various thicknesses. The insets show the enlarged part near the superconducting transition.

Figure 5 The field dependent transport critical current density $J_c$ of seven-core tapes (thickness = 1.00, 0.80 and 0.60 mm) and single-core tape (thickness = 0.60 mm).

Figure 6 SEM image of the cross sectional view of the Sr-122 core in (a) single-core and (b) seven-core tapes (rolling thickness = 0.60 mm).

Table 1

| Deformation process | Nominal Diameter / Thickness (mm) | $S_{Fe+Ag+Sr122}$ (mm$^2$) | $S_{Sr-122}$ (mm$^2$) | $S_{Sr122}/S_{Fe+Ag+Sr122}$ (%) | $T_{c, onset}$ (K) | $\rho_{300}$ ($\mu\Omega$cm) | Transport $J_c$ (kA/ cm$^2$) (at 4.2 K, 0 T) |
|---|---|---|---|---|---|---|---|
| Drawing | 2.00 | 3.299 | 0.290 | 8.79 | 33.5 | 14.5 | 4.4 |
| | 1.90 | 2.716 | 0.243 | 8.95 | 33.7 | 13.9 | 4.1 |
| | 1.80 | 2.487 | 0.208 | 8.36 | 33.8 | 12.8 | 3.6 |
| | 1.70 | 2.216 | 0.187 | 8.44 | 33.6 | 10.7 | 4.4 |
| Rolling | 1.50 | 3.153 | 0.250 | 7.93 | 33.1 | 13.1 | 2.2 |
| | 1.25 | 3.149 | 0.249 | 7.91 | 33.4 | 12.4 | 2.7 |
| | 1.00 | 2.991 | 0.243 | 8.12 | 33.3 | 11.7 | 2.7 |
| | 0.80 | 2.863 | 0.227 | 7.93 | 34.3 | 11.4 | 8.5 |
| | 0.60 | 2.587 | 0.209 | 8.08 | 34.5 | 10.1 | 21.1 |

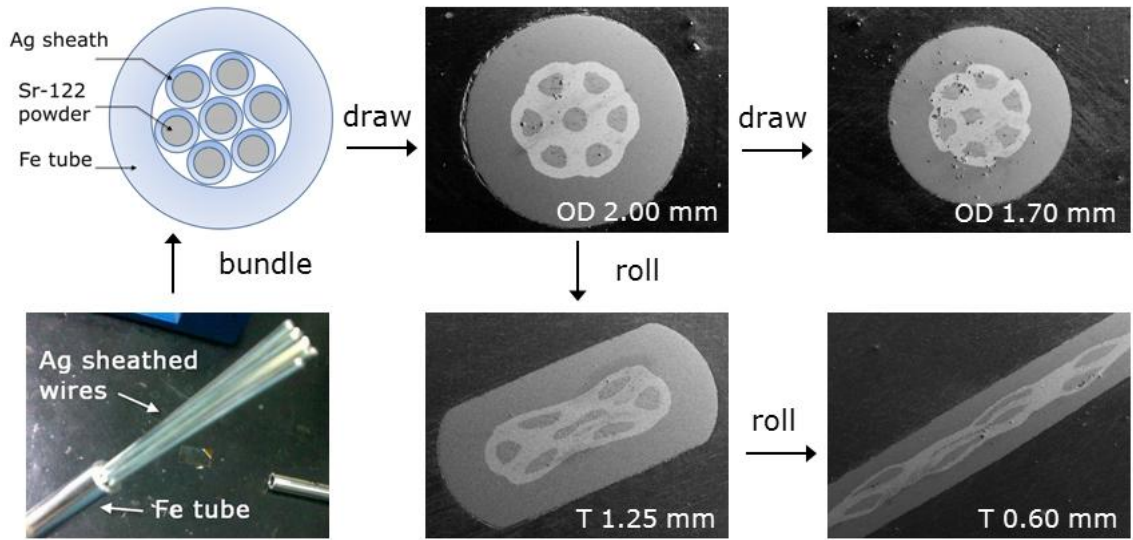

Figure 1 Yao et al.

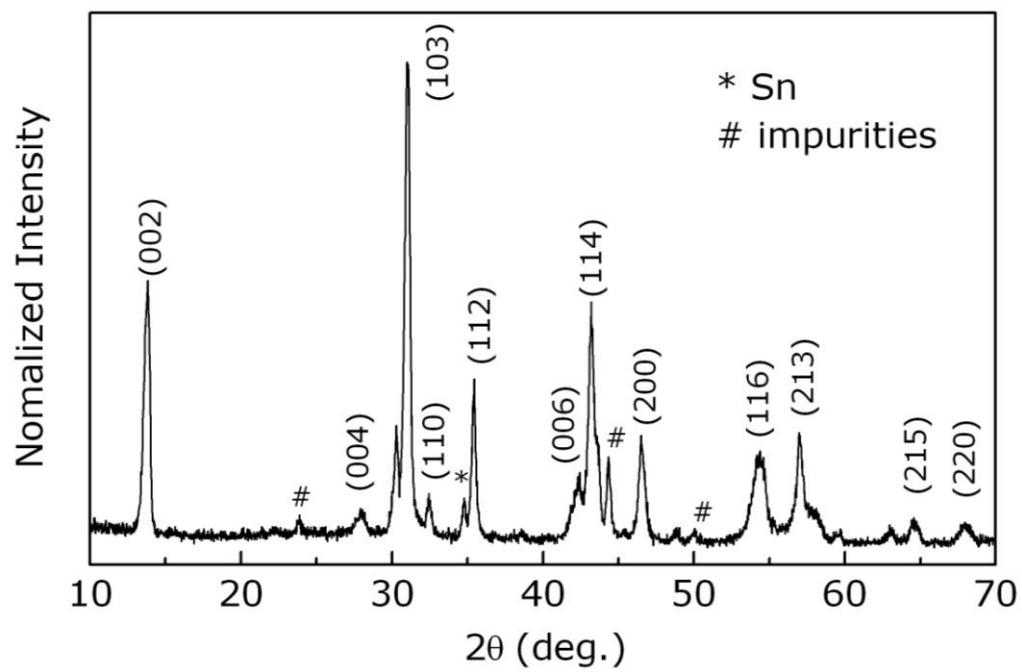

Figure 2 Yao et al.

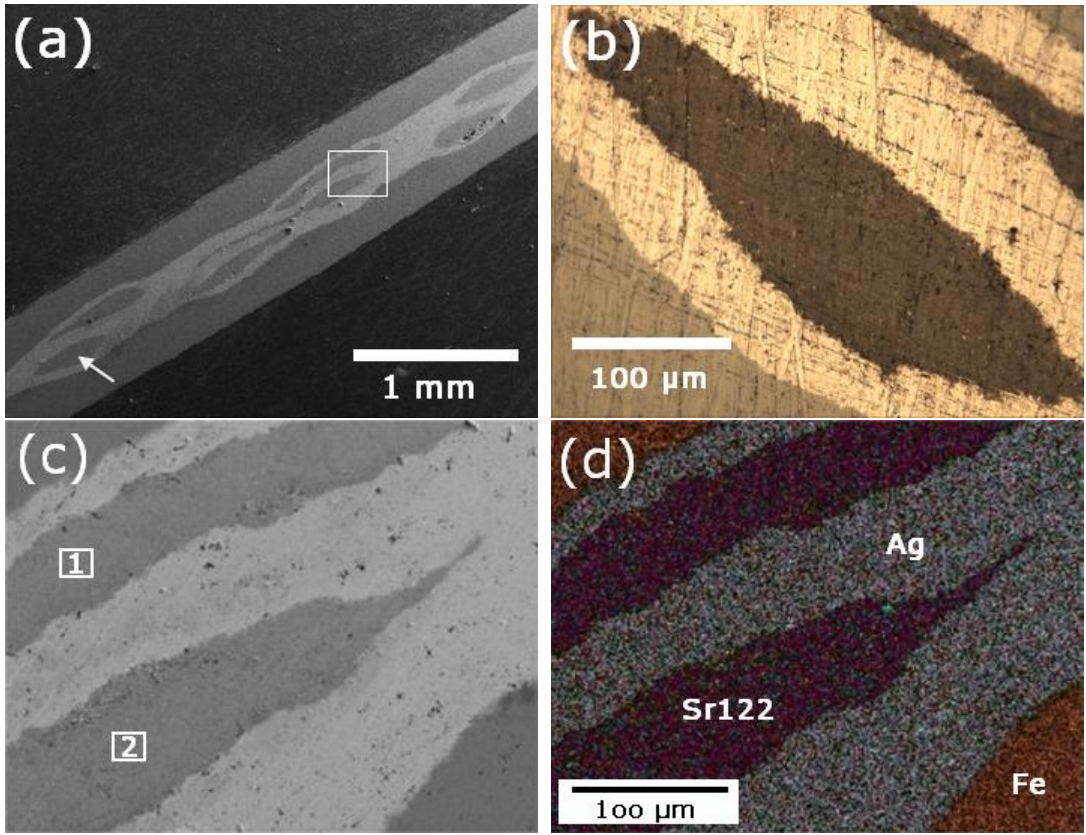

Figure 3 Yao et al.

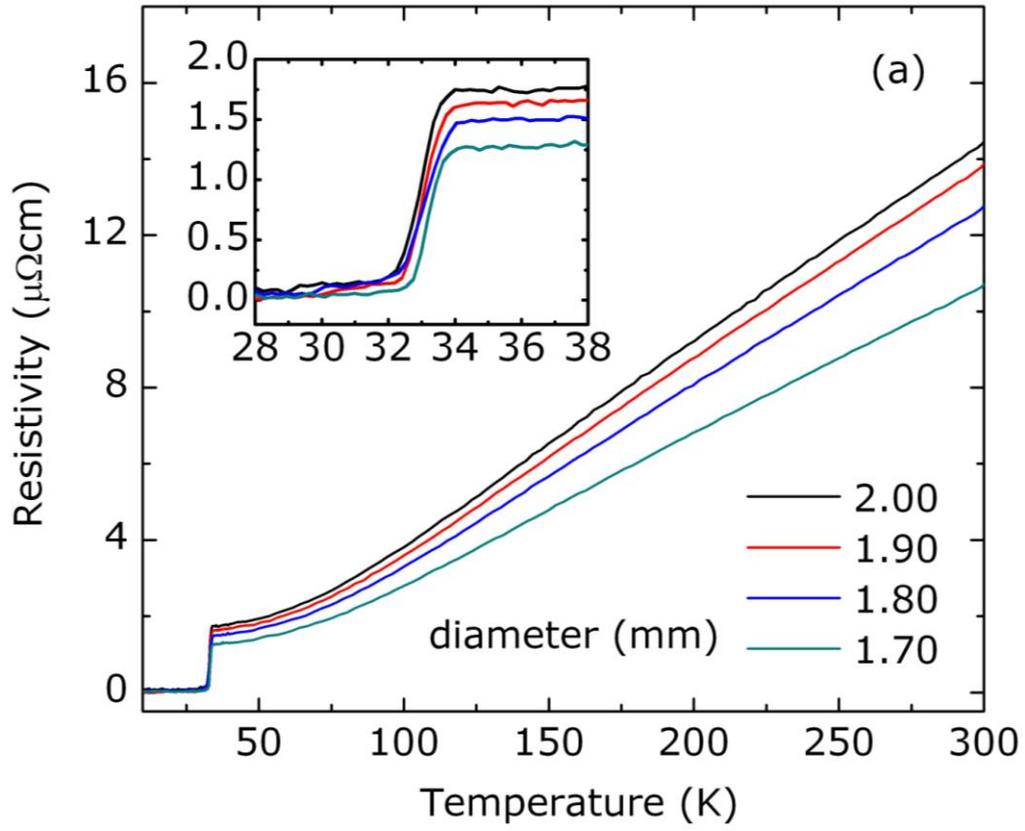

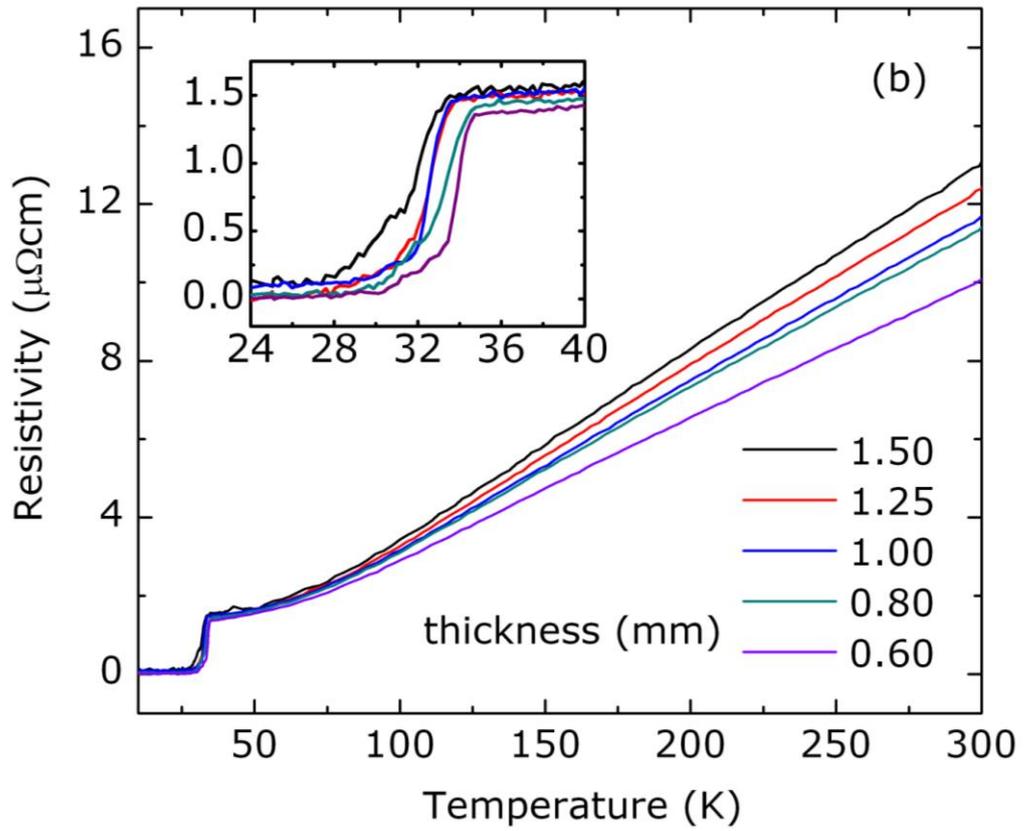

Figure 4 Yao et al.

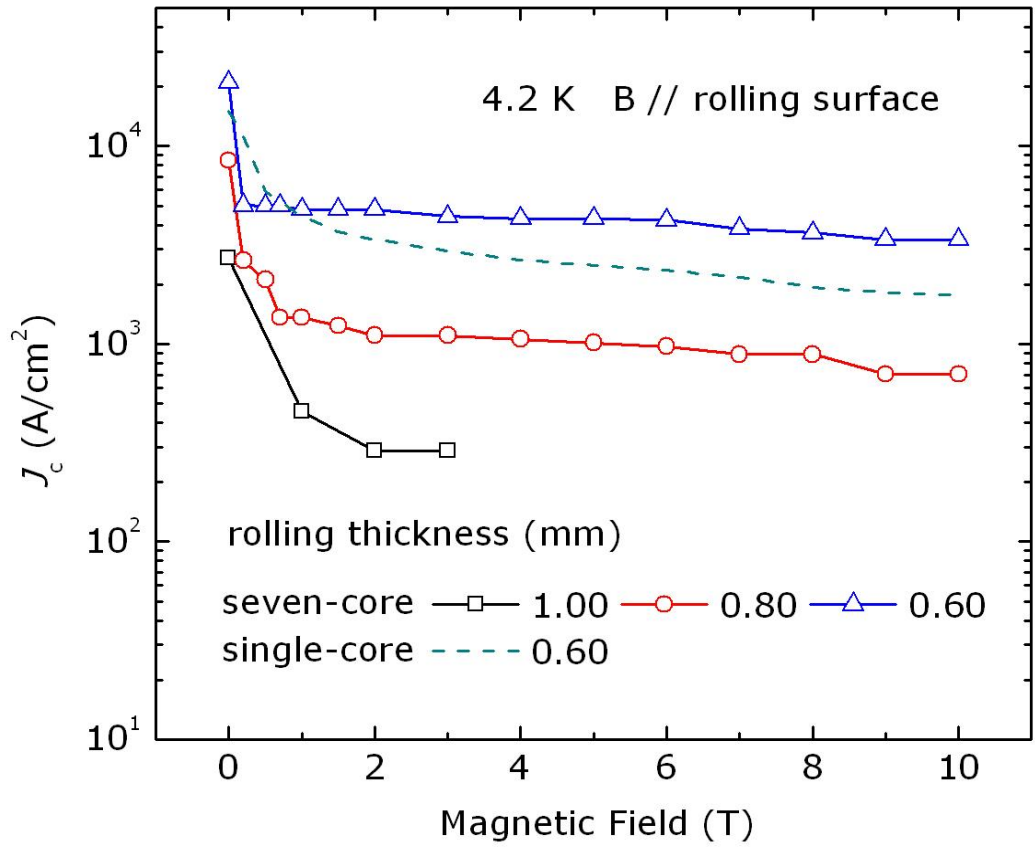

Figure 5 Yao et al.

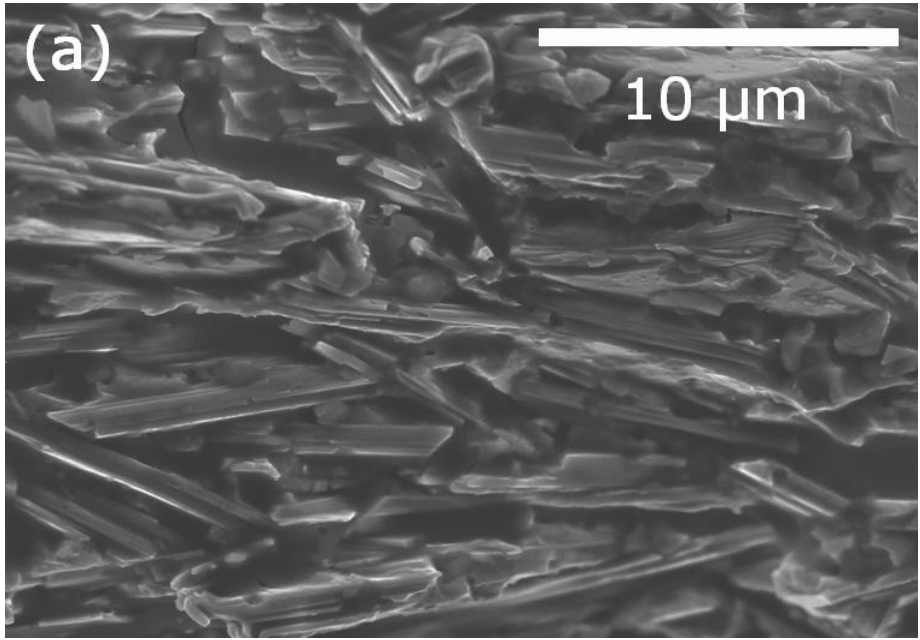
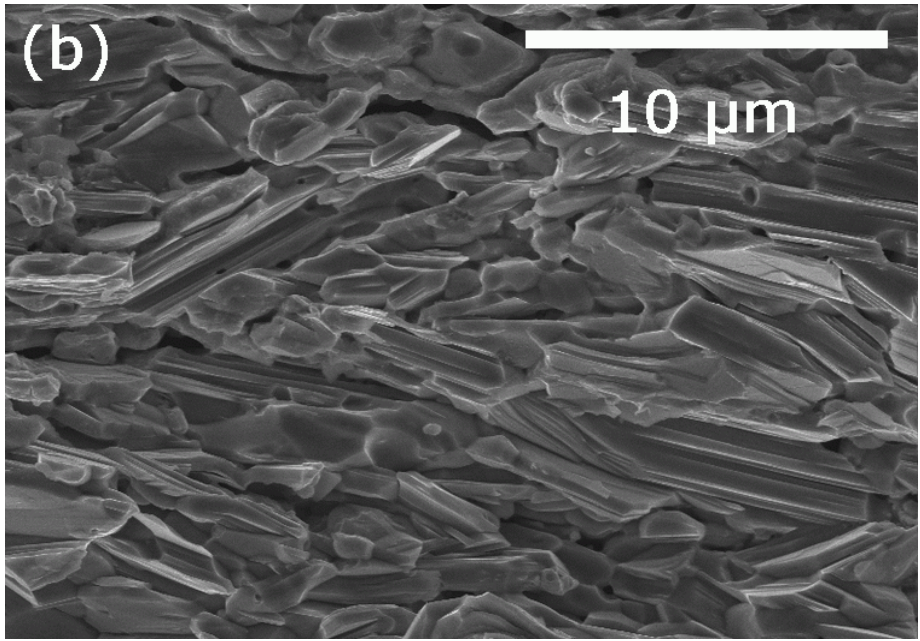

Figure 6 Yao et al.